\documentclass[journal]{IEEEtran}
\usepackage{amsfonts,amssymb}
\usepackage{amsmath}
\usepackage{amsthm}

\usepackage{graphicx}
\usepackage{enumerate}
\usepackage{multirow}
\usepackage{longtable}
\usepackage{rotating}
\usepackage{calc}
\usepackage{subfig}
\usepackage{graphicx}
\usepackage{epstopdf}
\usepackage{graphics}
\usepackage{epsfig}
\usepackage{psfrag}
\usepackage{cases}

\usepackage{array,color}
\usepackage{tabularx}
\usepackage{multirow}
\usepackage{booktabs}
\usepackage{makecell}

\usepackage{algorithm}

\usepackage{algorithmic}

\usepackage{multirow}

\usepackage[numbers,sort&compress]{natbib}
\usepackage{url}
\usepackage{bm}
\usepackage{amsfonts}
\usepackage{amsmath}
\usepackage{amssymb}
\usepackage{amsthm}
\usepackage{color}

\usepackage{array,color}%
\usepackage{tabularx}%
\usepackage{multirow}%
\usepackage{booktabs}%
\usepackage{makecell}  

\usepackage{graphicx}  
\usepackage{subfig}
\usepackage{epstopdf}
\usepackage{graphics}
\usepackage{epsfig}
\usepackage{psfrag}
\usepackage{overpic}

 \usepackage{stfloats}
 \usepackage{float}

 \usepackage{indentfirst} %

\usepackage{gensymb}

\begin{document}

\title{A New Approach of Exploiting Self-Adjoint Matrix Polynomials of Large Random Matrices for Anomaly Detection and Fault Location}
\author{Zenan Ling$^1$,Robert C. Qiu$^{1,2}$,~\IEEEmembership{IEEE Fellow}, Xing He$^1$, Lei Chu$^1$
\thanks{This work was partly supported by NSF of  China  No. 61571296 and (US)  NSF Grant No. CNS-1619250.

\textcircled{c} 20xx IEEE. Personal use of this material is permitted. Permission  from IEEE must be obtained for all other uses, in any current or future  media, including reprinting/republishing this material for advertising or  promotional purposes, creating new collective works, for resale or  redistribution to servers or lists, or reuse of any copyrighted  component of this work in other works.

$^1$ Department of Electrical Engineering,
Center for Big Data and Artificial Intelligence, State Energy Smart Grid Research and Development Center, Shanghai Jiaotong
 University, Shanghai 200240, China. (e-mail: ling\_zenan@163.com; rcqiu@sjtu.edu.cn; hexing\_hx@126.com; leochu@sjtu.edu.cn).

$^2$ Department of Electrical and Computer Engineering,
Tennessee Technological University, Cookeville, TN 38505, USA. (e-mail:rqiu@tntech.edu).
}}
\maketitle
\IEEEpeerreviewmaketitle
\begin{abstract}
Synchronized measurements of a large power grid enable an unprecedented opportunity to study the spatial-temporal correlations. Statistical analytics for those massive datasets start with high-dimensional data matrices. Uncertainty is ubiquitous in a future's power grid. These data matrices are recognized as random matrices. This new point of view is fundamental in our theoretical analysis since true covariance matrices cannot be estimated accurately in a high-dimensional regime. As an alternative, we consider large-dimensional sample covariance matrices in the asymptotic regime to replace the true covariance matrices.
The self-adjoint polynomials of large-dimensional random matrices are studied as statistics for big data analytics. The calculation of the asymptotic spectrum distribution (ASD) for such a matrix polynomial is understandably challenging. This task is made possible by a recent breakthrough in free probability, an active research branch in random matrix theory. This is the very reason why the work of this paper is inspired initially.
The new approach is interesting in many aspects. The mathematical reason may be most critical. The real-world problems can be solved using this approach, however.
\end{abstract}


\section{Introduction}

Among challenges for big data analytics towards grid modernization, data-driven approach and data utilization are of great significance in power system operation in smart grids~\cite{bda2016tsg}.
Current power systems are huge in size and complex in topology. Model-based methods can not always meet the real-life needs when assumptions and simplifications are prerequisites for these mechanism models. Massive datasets are accessible, however, when the operation of the power system is  monitored by  a large number of sensors such as phase measuurment units (PMUs). For instance, China has deployed 1717 PMUs as of 2013 ~\cite{lu2015advancing} and there are about 500 PMUs installed by July 2012 in America~\cite{nuthalapati2015managing}.

The synchronized measurements of a large power grid enable the joint modeling of temporal statistical properties across the spatial nodes. The spatial-temporal couplings pose opportunities and challenges. Towards this goal, random matrix theory (RMT)  is used for data analysis first in~\cite{he2015arch}. RMT starts  in the early 20th century. Due to the increase in the dimensionality of collected datasets, RMT has been commonly used for considering problems regarding the behavior of eigenvalues of large dimensional random matrices  in physics, finance, wireless communication, etc~\cite{qiu2015smart,qiu2013bookcogsen}.

Due to the large size of datasets, randomness or uncertainty is at the heart of  data modeling and analysis in a complex, large power gird  when rapid fluctuations in voltages and currents are ubiquitous. Often, these fluctuations exhibit some certain (or deterministic) distribution properties~\cite {lim2016svd}.  Our approach exploits the massive datasets across the large grid that are distributed in both spatially and temporally. Random matrix theory (RMT) appears very natural for the problem at hand. In a random matrix of  size ${\mathbb{C}^{N \times T}},$ we use $N$ variables to represent the spatial nodes. For the $i$-th node where $i=1,...,N$,  there are $T$ observations to represent the temporal samples $t=1,...,T.$ When the number of nodes $N$ and data samples $T$ are large, very unique mathematical phenomenon occurs such that power mathematical tools such as free probability~\cite{qiu2015smart} can be exploited to develop big data analytics for joint spatial-temporal datasets. This is the central purpose of this paper.

Free probability is a powerful tool for solving random matrix problems, such as additive and multiplicative free convolution. Based on free probability theory, asymptotic limits of the testing functions (free self-adjoint matrix polynomials), can be obtained numerically through certain algorithms. Closed form expressions exist only for some simple matrix polynomials. The obtained asymptotic limits provide the rigorous bounds in mathematics which can  help distinguish signals and noise in grid data. The \textit{anomaly detection} is conducted through hypothesis testing and an indicator for \textit{fault location} is designed using some mathematical tricks.
\subsection{Contributions of Our Paper}
This paper is built upon our previous work~\cite{he2015arch,he2015corr,He2016Designing,Chu2016Massive} in the last several years.  Motivated for machine learning from massive datasets, our line of research is based on the modern high-dimensional statistics where RMT is central to this paradigm. The contributions of this paper can be summarized as follows:
\begin{enumerate}
    \item  The aim of this paper is to exploit the polynomials of large random matrices in the context of big data analytics for a large power grid. To our best knowledge, this attempt is for the first time. This analysis is made possible by a \textit{recent breakthrough}~\cite{Belinschi2013Analytic,Speicher2015Polynomials} in the literature of mathematics. Our result represents one of the first applications of these algorithms in engineering.
	\item  Using the new analytic tool~\cite{Belinschi2013Analytic,Speicher2015Polynomials} from free probability theory, we are able to distinguish the noise and signals from grid data with provable mathematics guarantees. It is natural to conduct anomaly detection by hypothesis testing.
    \item Both linear and nonlinear polynomials of large random matrices can be handled in this new framework~\cite{Belinschi2013Analytic,Speicher2015Polynomials}. Simulations demonstrate that compared with the linear cases, nonlinear cases perform better in reducing  the false alarm probability.
    \item  Based on certain new algorithms developed in this paper,  an indicator for fault location is proposed and validated to be valid by simulations and real-world cases.
	\end{enumerate}
\subsection{Related Work}
There are numerous researches on data driven methods for modeling and analysing large power systems. Le, Chen and Kumar~\cite{chen2013dimensionality} propose a linearized analysis method  for early event detection using partial least squares estimation. Lim and DeMarco~\cite{lim2016svd,Lim2013Model} propose a singular value decomposition (SVD)-based voltage stability assessment  from principal component analysis (PCA). Both methods in the above are  PCA related and the selection of the eigenvalues has a crucial influence.  Lim et al select the largest eigenvalue and Xie et al~\cite{xie2014dimensionality} adopt  the method of the threshold of cumulative variance proportion. Their common disadvantage  is that the  pre-defined threshold depends on the experience without the consideration of the statistical characteristic of the grid data, so the redundancy or loss of information is unavoidable.

Also, along with the  new wave of deep learning, some Neural Network based methods are proposed. With powerful modeling ability of  neural networks, Eltigani~\cite{Eltigani2013Implementation} realize assessing the transient stability.  Zhou~\cite{Zhou2010Online}  present a method for long-term voltage stability monitoring based on Artificial Neural Network which requires training before online deployment. However, the training speed of networks slows down with the scale-up of the system and increase of training samples. Moreover, the high quality of the sampling data is crucial for the neural network's  generalization ability  which is not practical in current power system.

This paper is organized as follows. Section~\ref{section2} establishes the random matrix model for the power grid as the basis of this paper. Section~\ref{section3} proposes the anomaly detection method from the way of hypothesis testing and  an indicator for fault location. Section~\ref{section4} provides a brief introduction of  the algorithm for obtaining the asymptotic spectral distribution of free self-adjoint polynomial which is essential for our analytical framework. In Section~\ref{CASE STUDIES}, numerical case studies validate our methods  with simulation data and real-world data. In Section~\ref{signalest}, the estimation of signal strength is discussed under the linear assumption. Conclusion and further direction of this research are given in Section~\ref{CONCLUSION}. For the sake of simplicity, some details and the supplementary materials are deferred to the Appendix.

\section{Random Matrix Model and Data processing  for power grid}
\label{section2}

\subsection{Random Matrix Model for Power Grid}
\label{A1}
Following~\cite{he2015arch,He2016Designing}, the power flow equations, which define the equilibrium operating condition of a  power system, can be written as:
 \begin{equation}
\label{eq:powerflowquation1}
   \left[ {\begin{array}{*{20}{c}}
{\Delta P}\\
{\Delta Q}
\end{array}} \right] = J\left[ {\begin{array}{*{20}{c}}
{\Delta \theta }\\
{\Delta V}
\end{array}} \right] = \left[ {\begin{array}{*{20}{c}}
{\frac{{\partial P(\theta ,V)}}{{\partial \theta }}}&{\frac{{\partial P(\theta ,V)}}{{\partial V}}}\\
{\frac{{\partial Q(\theta ,V)}}{{\partial \theta }}}&{\frac{{\partial Q(\theta ,V)}}{{\partial V}}}
\end{array}} \right]\left[ {\begin{array}{*{20}{c}}
{\Delta \theta }\\
{\Delta V}
\end{array}} \right]
\end{equation}
where $P, Q, V, \theta$ denotes the active power, the reactive power, the voltage phase angle and the voltage amplitude  respectively.

To characterize the role of each block of the Jacobian matrix, denote:
\begin{equation}
\label{eq:powerflowquation2}
   \begin{array}{l}
H = \frac{{\partial P(\theta ,V)}}{{\partial \theta }},N = \frac{{\partial P(\theta ,V)}}{{\partial V}}\\
K = \frac{{\partial Q(\theta ,V)}}{{\partial \theta }},L = \frac{{\partial Q(\theta ,V)}}{{\partial V}}
\end{array}
\end{equation}

Then, taking the inverse of the Jacobian matrix $J$ in~~\eqref{eq:powerflowquation1} leads to~\eqref{eq:powerflowquation3}, providing the desired input-output relationship,
\begin{equation}
\label{eq:powerflowquation3}
   \left[ {\begin{array}{*{20}{c}}
{\Delta \theta }\\
{\Delta V}
\end{array}} \right] = \left[ {\begin{array}{*{20}{c}}
M&{ - MN{L^{ - 1}}}\\
{ - {L^{ - 1}}KM}&{{L^{ - 1}} + {L^{ - 1}}KMN{L^{ - 1}}}
\end{array}} \right]\left[ {\begin{array}{*{20}{c}}
{\Delta P}\\
{\Delta Q}
\end{array}} \right]
\end{equation}
where $M = {(H - N{L^{ - 1}}K)^{ - 1}}$.

Therefore, under the situation that $Q$ is relatively  constant, the model between $V$ and $P$ is obtained as:
\begin{equation}
\label{eq:RMTform1}
	\Delta V = \Xi \Delta P
\end{equation}
with $\Xi =  - {L^{ - 1}}KM$.

Considering   $T$ random vectors observed at time  $i=1,...,T,$ a random matrix is formed as follows:
 \begin{equation}
\label{eq:RMTform}
	\left[ {\Delta {{{{V}}_1}} , \cdots ,\Delta  {{{{V}}_T}}} \right] = \left[ { {{{\Xi}}_1\Delta{{P}}_1}, \cdots , {{{{\Xi}}_T\Delta {P}}_T}} \right].
\end{equation}

 It is worth noting that only voltage magnitude of PMU data is used. The voltage magnitude are more sensitive to topology change than phase angle and they  remain   relatively stable in normal operating condition~\cite{Lim2013Model}. Without dramatic topology changes, rich statistical empirical evidence indicates that the Jacobian matrix ${J}$  keeps nearly constant, so does $\Xi$. Thus  \eqref{eq:RMTform} is rewritten as:
\begin{equation}
\label{Eq:RMMVTP}
\mathbb{V} = {\Xi}{\mathbb{P}}_{N \times T}
\end{equation}
where $\mathbb{V} = \left[ {\Delta {{{{V}}_1}} , \cdots ,\Delta  {{{{V}}_T}}} \right]$, $ { {\Xi}}={\Xi}_1=\cdots={\Xi}_T,$ and $\mathbb{P} = \left[ {\Delta {{{{P}}_1}} , \cdots ,\Delta  {{{{P}}_T}}} \right]$
. Here $\mathbb{V}$ and $\mathbb{P}$ are random matrices. To model the fast time scale stochastic variation in a load, we assume that $\mathbb{P}$ is a random matrix with Gaussian random variables as its entries, following~\cite{he2015arch,He2016Designing}.

\subsection{Data Processing Method }
\label{1.2}
 The sampling data matrix  $\mathbb{V}$ of real power grid  is always non-Gaussian, so  a  normalization procedure in~\cite{he2015arch} is adopted to conduct data preprocessing. Meanwhile,  a Monte Carlo method is employed to estimate the empirical spectral distribution (ESD) of raw grid  data  according to the asymptotic property theory.

 The data processing procedure above is organized as following steps  in Algorithm~\ref{alg1}. The parameter $N$ denotes the number of buses and $T$ denotes the  sampling period. Note that  $\eta$ is extremely small, e.g. $\eta=10^{-5}$ and $M$ is set to 10 in our simulation cases.

 \begin{algorithm}[h]

    \caption{}
    \label{alg1}
    \begin{algorithmic}[1]
\REQUIRE ~~\\
The sample data matrices: ${\mathbb{V}}$;\\
The number of repetition times: $M$ (10 is enough);\\
The size of ${\mathbb{V}}$: $N,T$;\\
The variance of the small white noise ${\varepsilon}_{N\times T}$: $\eta$;\\
        \FOR{$i \leq M$}

            \STATE Add  small white noises ${\varepsilon}_{N\times T}$ to the sample data matrix\\
            $ \widetilde{\mathbb{V}}= \mathbb{V} +  {\varepsilon}_{N\times T}$ ;\\
            \STATE Standardize $\widetilde{\mathbb{V}}$ , i.e. mean=0, variance=1;
            \STATE Calculate the sample covariance matrices: ${\Sigma}=\widetilde{\mathbb{V}}\widetilde{\mathbb{V}}^{'}/T$;\\
            \STATE Calculate the eigenvalues of $\Sigma$;
        \ENDFOR
\STATE Calculate the  empirical spectral distribution of $\Sigma$;
\ENSURE ~~\\
The histogram of the ESD of $\Sigma$ .
    \end{algorithmic}
\end{algorithm}

\subsection{Validation of Proposed Model}
\label{1.3}
Marchenko-Pastur Law (M-P Law)~\cite{Marchenko1967Distribution}, a basic theorem in random matrix theory, is introduced  to verify the random matrix model for power grid.

\newtheorem{theorem}{\textbf{Theorem}}[section]
\begin{theorem}[M-P Law~\cite{Marchenko1967Distribution}]
 Let $X = \{ {x_{i,j}}\} $ be a $N \times T$ random matrix whose entries with the mean $\mu= 0$ and the variance ${\sigma ^2}<\infty $, are independent identically distributed (i.i.d). As $N, T \longrightarrow \infty$ with the ratio $ c=N/T \in (0,1] $.
 \begin{equation}
\label{defeq:SampleCov}
\Sigma = \frac{1}{T}X{X^H} \in {{\mathbb{C}}^{N \times N}}
 \end{equation}
	is the  corresponding sample covariance matrix. Then,  the asymptotic spectral distribution of $\Sigma$ is given by:
 \begin{equation}
  {\mu {' }}(x)=
  \begin{cases}
  \frac{1}{{2\pi x{\sigma ^2}}}\sqrt {(b - x)(x - a)} &\mbox{if $a\leq x\leq b$}\\
  0 &\mbox{otherwise}
  \end{cases}
 \end{equation}
 where $a = {\sigma ^2}{(1 - \sqrt c )^2}$, $b = {\sigma ^2}{(1 + \sqrt c )^2}$. Here, $\Sigma$ is called Wishart matrix.
 \label{theorem1}
\end{theorem}
According to Algorithm~\ref{alg1}, we obtain the ESD of  the sample covariance matrix $\Sigma$ of  real-world datasets for 34 PMUs. The $\mathbb{V}$ is collected in normal operation.

\begin{figure}[htpb]
\centering
\begin{overpic}[scale=0.3
]{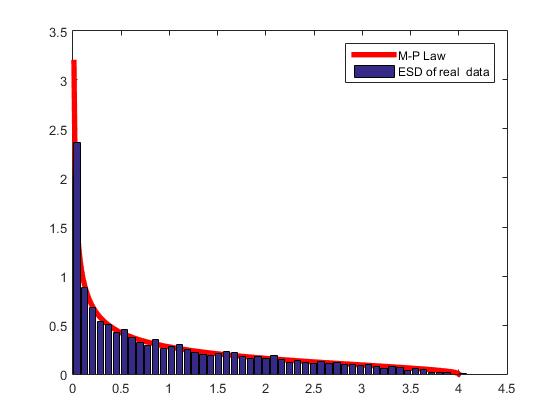}

\end{overpic}
\caption{Histogram of the empirical spectral distribution of the covariance of  34-PMU data collected in normal operation. The red curve represents the M-P Law.}
\label{fig:realdata}
\end{figure}

As illustrated in Fig~\ref{fig:realdata}, the histogram of the ESD of  $\Sigma$  coincides with  the M-P Law.  Although the asymptotic convergence is considered under infinite dimensions, i.e., $N \to \infty ,T \to \infty {\text{ but }}N/T \to c \in \left( {0,1} \right),$ the asymptotic results are fairly accurate for  moderate matrix sizes such as $N=10$s. It effectively explains why RMT is practical for the real-world datasets in a power grid.

\section{Anomaly Detection and Fault Location }
\label{section3}
\subsection{ Hypothesis Testing For Anomaly Detection}
\label{2.1}
Based on the random matrix model for power gird in ~\ref{A1}, the  problem of anomaly detection is formulated in terms of the hypothesis testing :
\begin{equation}
\label{model}
\left|
\begin{array}{*{20}{c}}
{{{\cal H}_0}:{\Sigma_1}  = {\Sigma_0}}\\
{{{\cal H}_1}:{\Sigma_1} \neq {\Sigma_0}}
\end{array}
\right.
\end{equation}
where ${\Sigma_0}$ is the sample covariance matrix of the grid data collected in normal operation and ${\Sigma_1}$ is the sample covariance matrix of the grid data for abnormal operation. This problem is a matrix hypothesis testing~\cite{qiu2015smart,qiu2013bookcogsen}. Test statistics are central to hypothesis testing.

In this paper $P(\Sigma_1,\Sigma_0)$ is adopted as test statistics. Here, $P$ is a self-adjoint polynomial of large random matrices, i.e. $P=P^H.$  The $P(\Sigma_1,\Sigma_0)$ measures the difference between two sample covariance matrices.

\newtheorem{theorem2}{\textbf{Theorem}}[section]
\begin{theorem}[The self-adjoint matrix polynomial of large Hermitian random matrices~\cite{Speicher2015Polynomials}]
  Let ${\bf \Sigma}_N= (\Sigma_1^{(N)},...,\Sigma_p^{(N)})$ be a family of independent, normalized $N\times N$ Wishart matrices. Assume that for every Hermitian matrix $P_N$ of the form
  \begin{equation}
  P_N=P({\bf \Sigma}_N)
\end{equation}
where P is a free self-adjoint matrix polynomial, we have with probability one that:
    1. The empirical spectral distribution  of a free self-adjoint matrix polynomial $P_N$ converges weakly to a compactly supported $\mu$ on the real line as $N$ goes to infinity.

    2. For any $\varepsilon >0$, almost surely there exits $N_0$ such that for all $N>N_0$, $Sp(P_N)\subset Supp(\mu)+(-\varepsilon ,\varepsilon)$, where 'Sp' means the spectrum and 'Supp' means the support.
\label{theorem2}
\end{theorem}
Theorem~\ref{theorem2} implies that if  ${\cal H}_0$ is true, the ESD of $P(\Sigma_1,\Sigma_0)$ will coincide with a theoretical curve\footnote{This curve can be calculated by a certain algorithm. See Section~\ref{section3} for details.}, i.e. the asymptotic spectral distribution (ASD) of $P$. Besides, no  eigenvalue exits outside of the support of the theoretical curve.

In this paper, the eigenvalues outside of the support are called \textit{outliers}.

According to Theorem~\ref{theorem2}, the proposed detection method is summarized as follows:
\begin{enumerate}
	\item Calculate ${\Sigma_0}$ and  ${\Sigma_1}$ from the sample data with the preprocessing method stated in Algorithm~\ref{alg1}.
	\item Compare the theoretical curves corresponding with the ESDs of different matrix polynomials $P(\Sigma_1,\Sigma_0)$.
	\item Anomaly detection is conducted: if \textit{outlier} exists, ${\cal H}_0$ will be rejected, i.e. signals exist in the system.
\end{enumerate}

Based on the hypothesis testing \eqref{model}, we propose a statistic indicator denoted by
\begin{equation}
\label{signalstrenth}
s = \frac{{\sum\limits_{{\lambda _k} \in {\rm{outliers}}} {{\lambda _k}} }}{{\sum\limits_{{\lambda _k} \notin {\rm{outliers}}} {{\lambda _k}} }}.
\end{equation}
The function of $s$ is similar to  the signal-to-noise ratio.

Notice that our proposed detection method is quite sensitive to the signal even if the signal is extremely weak~\cite{Loubaton2011Almost}. In order to reduce the false alarm probability,  it is necessary for the values of $s$  of the  normal  and the  abnormal load variation  to be different. So the choice of the polynomial functions is crucial.

In this paper, we study two typical self-adjoint matrix polynomials.
The first one is the multivariate\textit{ linear} polynomial:
\begin{equation}
\label{eq12}
{P_1}({\Sigma_0},{\Sigma_1}) = {\Sigma_1} - {\Sigma_0}.
\end{equation}

The second one is the  multivariate \textit{nonlinear} polynomial:
\begin{equation}
\label{eq13}
{P_2}({\Sigma_0},{\Sigma_1}) = ({\Sigma_1}-{\Sigma_0})^2 .
\end{equation}

Here, both ${\Sigma_0}$ and ${\Sigma_1}$ are the sample covariance matrices. The simulation results in Section~\ref{CASE STUDIES} will show that the performance of  the nonlinear polynomial is much better than the linear one.

It is difficult to obtain the ASD of free self-adjoint polynomials $P_1$ and $P_2$. Fortunately, the \textit{recent breakthrough}~\cite{Belinschi2013Analytic,Speicher2015Polynomials} in free probability in random matrix theory  has made this possible. To make the paper self-contained, the algorithm for calculating the ASD of $P$ is introduced briefly in Section~\ref{section3}.

\subsection{Fault Location}
\label{faultlocation}
 In this subsection, we investigate the fault location  based on the proposed anomaly detection method in~\ref{2.1}. Since the selected polynomials $P(\Sigma_0,\Sigma_1)$   are real and symmetric, the following equations
 \begin{equation}
P = v\left[ {\begin{array}{*{20}{c}}
{{\lambda _1}}&{}&{}\\
{}& \ddots &{}\\
{}&{}&{{\lambda _N}}
\end{array}} \right]u,
\end{equation}

\begin{equation}
P{v_k} = {\lambda _k}{v_k}
\label{eq14}
\end{equation}
hold. Here, $v$,$u$ denote the left and right eigenvector matrix; $\lambda_k$ is an eigenvalue of $P(\Sigma_0,\Sigma_1)$ and it indicates the energy of the corresponding eigenvector ${v_k}$.

For the element $P_{ij}$ in $P$, the derivative of \eqref{eq14} leads to the following:
\begin{equation}
\frac{{dP}}{{d{P_{ij}}}}{v_k} + P\frac{{d{v_k}}}{{d{P_{ij}}}} = \frac{{d{\lambda _k}}}{{d{P_{ij}}}}{v_k} + {\lambda _k}\frac{{d{v_k}}}{{d{P_{ij}}}}.
\label{eq15}
\end{equation}

 Left multiply \eqref{eq15} by $u_k^T$. Note that  ${u_k}^T{v_k} = 1$ and $u^T=v$ and we have
\begin{equation}
\frac{{d{\lambda _k}}}{{d{P_{ij}}}} = {u_k}^T\frac{{dP}}{{d{P_{ij}}}}{v_k}.
\label{eq16}
\end{equation}

Let  $\psi  = \frac{{dP}}{{d{P_{ij}}}}$. Obviously, only $\psi_{ij}=1$ and other elements of $\psi$ equal to zero. So \eqref{eq16} is simplified as:

\begin{equation}
\frac{{d{\lambda _k}}}{{d{P_{ij}}}} = {u_{kj}}{v_{ik}}.
\label{eq17}
\end{equation}

Finally, the contribution of the $i$-th row to the eigenvalue $\lambda_k$ is obtained by:
\begin{equation}
\sum\limits_{j = 1}^T {{{(\frac{{d{\lambda _k}}}{{d{P_{ij}}}})}^2}}  = \sum\limits_{j = 1}^T {{{({u_{kj}}{v_{ik}})}^2} = } {\rm{ }}{v_{ik}}^2\sum\limits_{j = 1}^T {{{({u_{kj}})}^2} = } {v_{ik}}^2.
\label{eq18}
\end{equation}

 In the work of Lim et al~\cite{Lim2013Model},    the singular vector correspondingto the largest singular value  is used  to conduct fault location. The simulation  results in~\cite{Lim2013Model} show  that the singular vector tells which buses are contributing to  the corresponding singular value.

 For the hypothesis testing in ~\ref{2.1}, not only the largest eigenvalue of the covairance matrix but also the  \textit{outliers} are viewed as the ``signals''. This observation inspires us to improve Lim's method by studying those eigenvectors corresponding to \textit{outliers}.
In particular, we design a  new location indicator denoted by
\begin{equation}
\label{indicator}
 {L_i} = \frac{{\sum\limits_{{\lambda _k} \in {\rm{outliers}}} {{\lambda _k}v_{ik}^2} }}{{\sum\limits_{{\lambda _k} \in {\rm{outliers}}} {{\lambda _k}} }},
\end{equation}
to quantify each bus's contribution to the anomaly.
Since that $v_{_{ik}}^2 \in [0,1]$ and $\sum\limits_i {v_{ik}^2}  = 1$,
 obviously,
 \begin{equation}
  \sum\limits_i {\sum\limits_{{\lambda _k} \in {\rm{outliers}}} {{\lambda _k}v_{ik}^2} }  = \sum\limits_{{\lambda _k} \in {\rm{outliers}}} {{\lambda _k}}
  \end{equation}
 Thus, $L_i \in (0,1]$ and $\sum\limits_i {L_{i}}= 1$. From the above, $L_i$ is a reasonable indicator that measures the  correlation between the $i$-th bus and the load variation.
The location (denoted as $loc$) of the most sensitive bus can be expressed as
\begin{equation}
\label{loc}
 loc = \underbrace {\arg \max }_{i \in \left( {1,...,N} \right)}{L_i}.
\end{equation}
The simulation results in \ref{case3} show that our method performs better than Lim's work~\cite{Lim2013Model} in the fault location task.

All the eigenvalues outside the support of the theoretical curve are considered.  The functions of $P(\Sigma_0,\Sigma_1)$ play a role of machine learning that classify between the noise and   signals. \textit{Outliers} are remained  as the useful signals for anomaly detection and fault location.

\section{The asymptotic spectral distribution of free selfadjiont  polynomial}
\label{section4}

In this section, we study  the asymptotic spectral distribution (ASD) of $P_1$ and $P_2$, on the premise that both ${\Sigma_0}$ and ${\Sigma_1}$ are Wishart matrices.
\subsection{The ASD of ${P_1}$}
\label{section3.1}

For obtaining the ASD of ${P_1}$, we introduce the operator-valued setting~\cite{Speicher2015Polynomials} briefly.
Let $\mathcal{A}$ be a unital algebra and $\mathcal{B}\subset \mathcal{A}$ be a subalgebra containing the unit. A linear map $E:\mathcal{A}\to \mathcal{B}$ is a conditional expectation.
For a random variable $x\in \mathcal{A}$, we define the operator-valued Cauchy transform: $G(b):=E[{{(b-x)}^{-1}}]   (b\in \mathcal{B})$ for which $(b-x)$ is invertible in $\mathcal{B}$.
Let ${{\mathbb{H}}^{+}}(\mathcal{B}):=\{b\in \mathcal{B}|\Im b>0\}$. In the following theorem \cite{Belinschi2013Analytic}, we will use the notation $h(b):=\frac{1}{G(b)}-b$.

\newtheorem{theorem2.1}{\textbf{Theorem}}[section]
 \begin{theorem}[\cite{Belinschi2013Analytic}]
 Let $x$  and $y$ be self-adjoint operator-valued random variables free over $\mathcal{B}$. Then there exists a Frechet analytic map
$\omega :{{\mathbb{H}}^{+}}(\mathcal{B} )\to {{\mathbb{H}}^{+}}(\mathcal{B})$ so that

$\bullet\Im {{\omega }_{j}}(b)\ge \Im b$ for all  $b\in {{\mathbb{H} }^{+}}(\mathcal{B} )$, $j\in \left\{ 1,2 \right\}$

$\bullet{{G}_{x}}({{\omega }_{1}}(b))={{G}_{y}}({{\omega }_{2}}(b))={{G}_{x+y}}(b)$

Moreover, if $b\in {{\mathbb{H}}^{+}}(\mathcal{B})$ , then ${{\omega }_{1}}(b)$  is the unique fixed point of the map.
${{f}_{b}}:{{\mathbb{H}}^{+}}(\mathcal{B} )\to {{\mathbb{H}}^{+}}(\mathcal{B} ),  {{f}_{b}}(\omega )={{h}_{y}}({{h}_{x}}(\omega )+b)+b, $
 and ${{\omega }_{1}}(b)\text{=}\underset{n\to \infty }{\mathop{\lim }}\,{{f}_{b}}^{on}(\omega )$ for any $\omega \in {{\mathbb{H}}^{+}}(\mathcal{B} )$, where $f_{b}^{on}$ means the n-fold composition of ${{f}_{b}}$ with itself. Same statements hold for ${{\omega }_{\text{2}}}(b)$, with replaced by $\omega \to {{h}_{x}}({{h}_{y}}(\omega )+b)+b. $
  \label{th1}
  \end{theorem}

\newtheorem{xxxx}{\textbf{Theorem}}[section]
 \begin{theorem}[Stieltjes inversion formula \cite{Pielaszkiewicz2015Closed}]
 For any open interval $I=(a,b)$ , such that neither a nor b are atoms for the probability measure $\mu $, the inversion formula
	\[\mu (I)=-\frac{1}{\pi }\int\limits_{I}{\Im ({{G}_{u}}(x+iy))dx}\]
holds.

 \end{theorem}

Theorem \ref{th1} provides an iterative algorithm to compute the operator-valued Cauchy transform of ${P_1}$. Then, the ASD  of ${P_1}$ is easily obtained through the Stieltjes inversion formula.

\subsection{The ASD of ${P_2}$}
\label{asd2}
  The ASD of ${P_2}$ is obtained  by linearizing the nonlinear polynomial.  Through Anderson's  linearation trick \cite{Anderson2011Convergence}, we have a
 procedure that leads finally to an operator:
  \[{{L}_{p}}={c}\otimes 1+{{b}_{0}}\otimes {{\Sigma}_{0}}+\cdots {{b}_{n}}\otimes {{\Sigma}_{n}}.\]
 In the case of ${P_2}$,  \[{L_{{P_2}}} = \left( {\begin{array}{*{20}{c}}
0&{{\Sigma_1-\Sigma_0}}&{{\frac{{{\Sigma _1} - {\Sigma _0}}}{2}}}\\
{{\Sigma_1-\Sigma_0}}&0&{ - 1}\\
{{\frac{{{\Sigma _1} - {\Sigma _0}}}{2}}}&{ - 1}&0
\end{array}} \right)\]

Therefore, ${L_{{P_2}}}$ can be easily written in the form of ${L_{{P_2}}}={c}\otimes 1+{{b}_{0}}\otimes {{\Sigma}_{0}}+ {{b}_{1}}\otimes {{\Sigma}_{1}}$, where ${c} = \left( {\begin{array}{*{20}{c}}
0&0&0\\
0&0&{ - 1}\\
0&{ - 1}&0
\end{array}} \right),$
${b_0} = \left( {\begin{array}{*{20}{c}}
0&-1&-\frac{1}{2}\\
-1&0&0\\
-\frac{1}{2}&0&0
\end{array}} \right),$
${b_1} = \left( {\begin{array}{*{20}{c}}
0&1&\frac{1}{2}\\
1&0&0\\
\frac{1}{2}&0&0
\end{array}} \right).$

\newtheorem{theorem3.1}{\textbf{Theorem}}[section]
\label{theorem3.1}
\begin{theorem}[\cite{Belinschi2013Analytic}]
 Consider that $p\in \mathbb{C}<{{X}_{1}},\ldots ,{{X}_{n}}>$ that has a self-adjoint linearization\[{{L}_{p}}={{b}_{0}}\otimes 1+{{b}_{1}}\otimes {{X}_{1}}+\cdots {{b}_{n}}\otimes {{X}_{n}}.\]
 Let $$\Lambda (z):=\left[ \begin{matrix}
   z & 0 & \cdots  & 0  \\
   0 & 0 & \cdots  & 0  \\
   \vdots  & \vdots  & \ddots  & \vdots   \\
   0 & 0 & \cdots  & 0  \\
\end{matrix} \right]     \quad    for \;all\; {z}\in {\mathbb{C}}.$$

Then, for each $z\in {{\mathbb{C}}^{+}}$ and all  small enough $\varepsilon >0$ , the operators $z-P\in \mathcal{A}$ and ${{\Lambda }_{\varepsilon }}(z)-{{L}_{P}}\in {{M}_{N}}(\mathbb{C})\otimes \mathcal{A}$ are both invertible and  \[{{G}_{P}}(z)\text{=}\underset{\varepsilon \to \text{0}}{\mathop{\lim }}\,{{\left[ G{}_{{{L}_{P}}}({{\Lambda }_{\varepsilon }}(z)) \right]}_{1,1}}\quad for\; all \;z\in {{\mathbb{C}}^{+}}\] holds.

\label{co1}
\end{theorem}
Theorem~\ref{theorem3.1} illustrates the relationship between the Cauchy transform of $P_2$ and $L_{P_2}$. Thus,  the nonlinear problem is turned  into the linear problem which is solved in ~\ref{section3.1}.

Finally, the ASD of ${P_2}$ is obtained through the application of Theorem \ref{th1} and the Stieltjes inversion formula. See  specific procedures of the algorithm in Appendix~\ref{appendixA}.

\section{CASE STUDIES}
\label{CASE STUDIES}
The proposed method is tested with the simulated data generated from IEEE 118-bus and a Polish 2383-bus system, respectively. Detailed information of the system is referred to the case118.m and case118.m in Matpower package and Matpower 4.1 User's Manual \cite{Zimmerman2011MATPOWER}.

 In subsection~\ref{case1},~\ref{case2} and~\ref{case3}, the proposed method is tested with simulated data in the standard IEEE 118-bus system, as shown in Fig. \ref{fig:IEEE118network} in Appendix~\ref{appendixB}.  The results generated form the Polish 2383-bus system are deferred to the Appendix~\ref{appendixC}.

 In subsection~\ref{case4},  the fault location method is validated by  real-world 34-PMU data.

For Cases 1-3, set the sample dimension, i.e. the number of buses, as $N=118$.
The spectral density distribution of free adjoint polynomial can be obtained through the algorithm in  Section~\ref{section3} as long as $N\le T$. Meanwhile, the sample dimension $N$ is required to be large enough to guarantee the accuracy of results in the proposed asymptotic theory of eigenvalue distributions. Therefore, in Cases 1-3 , we set the sample length to be equal to $N$, i.e. $T=118$, $c=T/N=1$ and select six sample voltage matrices presented in Tab. \ref{tab1}. The load variation is shown in Fig. \ref{fig:loadevent} in  Appendix~\ref{appendixB}.

\begin{table}[h]

\centering

\caption{System status and sampling data}

\begin{tabular}{p{2.05cm}|p{2.2cm}|p{3.52cm}}
\hline
  Cross Section (s)& Sampling (s) &Descripiton\\
\hline
  $\textbf{C}_0:118-900$ & $V_0:100\sim217$&Reference, no signal\\
  $\textbf{C}_1:901-1017$ & $V_1:850 \sim 967$&Existence of a step signal for Bus 22\\
  $\textbf{C}_2:1918-2600$ & $V_2:2200 \sim 2317$&Steady load growth for Bus 22\\
  $\textbf{C}_3:3118-3790$ & $V_3:3300 \sim 3417$&Steady load growth for Bus 52\\
  $\textbf{C}_4:3908-4100$ & $V_4:3900 \sim 4017$&Chaos due to voltage collapse\\
  $\textbf{C}_5:4118-5500$ & $V_5:4400 \sim 4517$&No signal\\
\hline
\end{tabular}
\label{tab1}
\raggedright
 {*We choose the temporal end edge of the sampling matrix as the marked time for the cross section. E.g., for $V_0:100\sim217$, the temporal label is 217 which belong to $\textbf{C}_0:118-800$.}
 \end{table}

 Power grid operates with only white noises during 0 s to 900 s; we choose sampling matrix {$V_0$}  as  the reference. Similarly, we mark other kinds of system operation status as  $\textbf{C}_1$--$\textbf{C}_5$, and choose their relevant sampling matrix {$V_1$}--{$V_5$} for the test.

\subsection{Case 1: Anomaly Detection with the Multivariate Linear Polynomial $P_1$}

\label{case1}

\begin{figure}[htbp]
 \subfloat[Step signal $V_1$]{\label{fig1a}
 \includegraphics[width=0.25\textwidth]{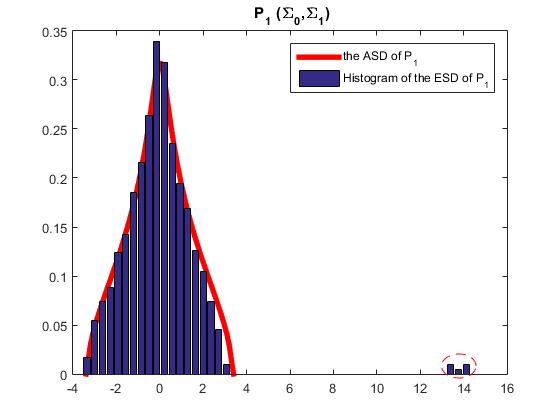}
 \label{fig2.a}}
 \subfloat[Stable growth A $V_2$]{\label{fig1b}
 \includegraphics[width=0.25\textwidth]{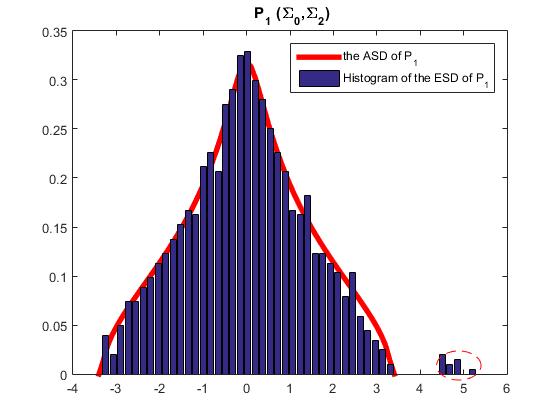}
 \label{fig2.b}}\\
 \subfloat[Stable growth B $V_3$]{\label{fig1c}
 \includegraphics[width=0.25\textwidth]{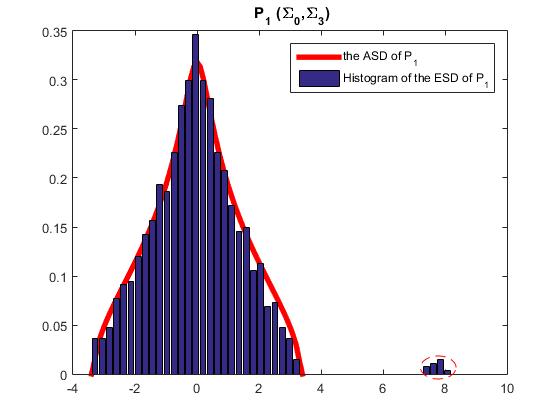}
 \label{fig2.c}}
 \subfloat[Voltage collapse $V_4$]{\label{fig1d}
 \includegraphics[width=0.25\textwidth]{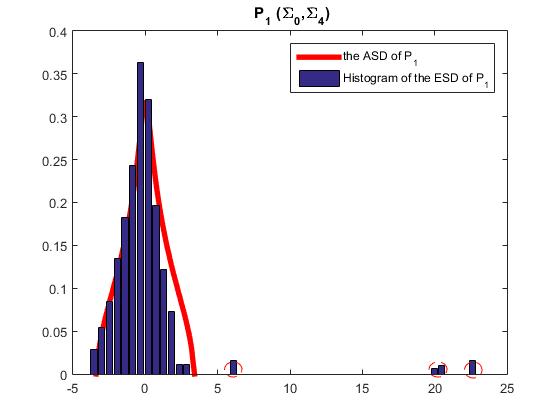}
 \label{fig2.d}}\\
 \subfloat[White noises $V_5$]{\label{fig1e}
 \includegraphics[width=0.25\textwidth]{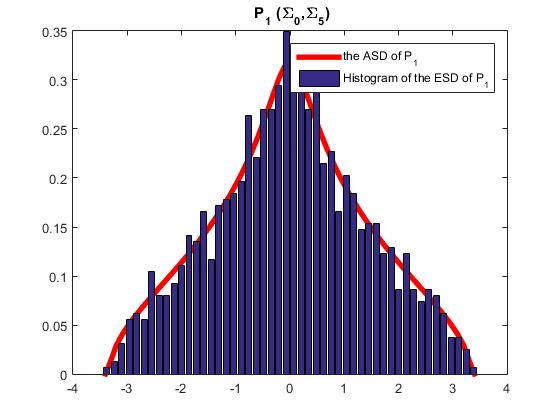}
\label{fig2.e} }
 \caption{Detection results with the multivariate linear polynomial $P_1$}
\label{result_p1}
 \end{figure}

We conduct anomaly detection using ${V_i}$ $(i \ge 1)$  and  the  reference matrix {$V_0$} through the proposed hypothesis testing \eqref{model}. Note that covariance matrix {$\Sigma_i$} is generated from ${V_i}$ by the preprocessing in ~\ref{1.2}.
Then, we choose the multivariate linear polynomial
 \[{P_1}({\Sigma_0},{\Sigma_i}) = {\Sigma_i} - {\Sigma_0},~i=1,2,3,4,5\]
to conduct this detection.
 Simulation results are shown in Fig.~\ref{result_p1}. The red curve represents  the ASD of ${P_1}$ obtained by Theorem~\ref{th1}. The  ESD histogram of ${P_1}({\Sigma_0},{\Sigma_i})$ is plotted by using Algorithm~\ref{alg1}. \textit{Outliers} are highlighted by ellipses. The values of $s$ defined in ~\eqref{signalstrenth} in $\textbf{C}_1$--$\textbf{C}_5$ are presented in Tab.~\ref{tabs1}.

1) For Fig.~\ref{result_p1}(a)-(d), the actual histograms agree with the theoretical curve very well except a few spikes (called outliers); (e): for the white noise case, there are no \textit{outliers}, as expected by the theory.

2) Sort by the values of $s$: $\textbf{C}_5<\textbf{C}_2<\textbf{C}_3<\textbf{C}_1<\textbf{C}_4$.

From result 1),  we observe that this method distinguishes the signals from  the white noise successfully. Result 2) implies that the size of \textit{outliers}  indicates the  strength of the signals.

However, the discrimination between the ramp signals and the step signals are  not very obvious. See Table~\ref{tabs1} for details.  To deal with this disadvantage, a direct approach is to increase the sensitivity for a given signal-to-noise ratio. In particular, a nonlinear polynomial $P_2$ is adopted.

\begin{table}[h]

\centering

\caption{The values of $s$ in $\textbf{C}_1$--$\textbf{C}_5$}

\begin{tabular}{p{1.5cm}|p{3.4cm}|p{0.9cm}}
\hline
  Cross Section & Description &$s$\\
\hline
  $\textbf{C}_1$ & Existence of a step signal &0.1096\\
  $\textbf{C}_2$ & Steady load growth &0.0357\\
  $\textbf{C}_3$ & Steady load growth &0.0609\\
  $\textbf{C}_4$ & Chaos due to voltage collapse &0.4219\\
  $\textbf{C}_5$ & No signa &0.000\\
\hline
\end{tabular}
\label{tabs1}
 \end{table}

\subsection{Case 2: Anomaly Detection with  the Multivariate Nonlinear Polynomial $P_2$}
\label{case2}

In Case 2, the process  is similar to \ref{case1} except that the test statistic function is replaced with \[{P_2}({\Sigma_0},{\Sigma_i}) = ({\Sigma_i}-{\Sigma_0})^2,~i=1,2,3,4,5\]
which is  a multivariate  nonlinear polynomial in two matrices. This is the simplest second-order polynomial. Of course, we can study higher orders that will be left for the work in a next paper.

\begin{figure}[htbp]
 \subfloat[Step signal $V_1$]{\label{fig3a}
 \includegraphics[width=0.25\textwidth]{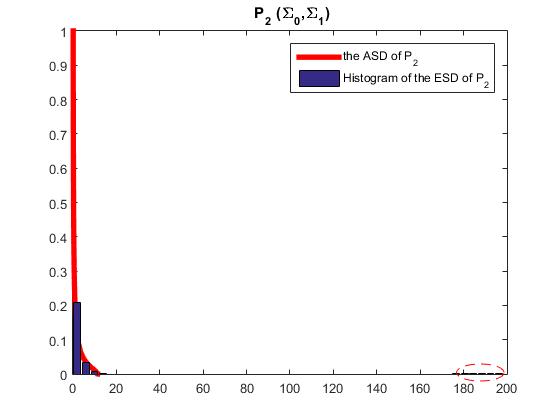}
 }
 \subfloat[Stable growth A $V_2$]{\label{fig3b}
 \includegraphics[width=0.25\textwidth]{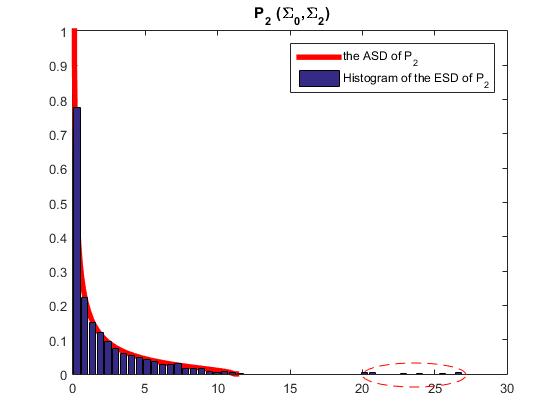}
 }\\
 \subfloat[Stable growth B $V_3$]{\label{fig3c}
 \includegraphics[width=0.25\textwidth]{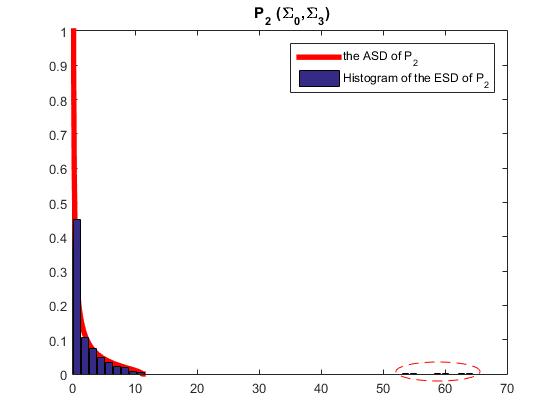}
 }
 \subfloat[Voltage collapse $V_4$]{\label{fig3d}
 \includegraphics[width=0.25\textwidth]{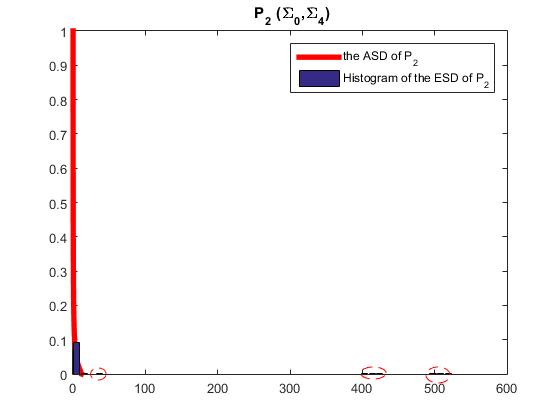}
 }\\
 \subfloat[White noises $V_5$]{\label{fig3e}
 \includegraphics[width=0.25\textwidth]{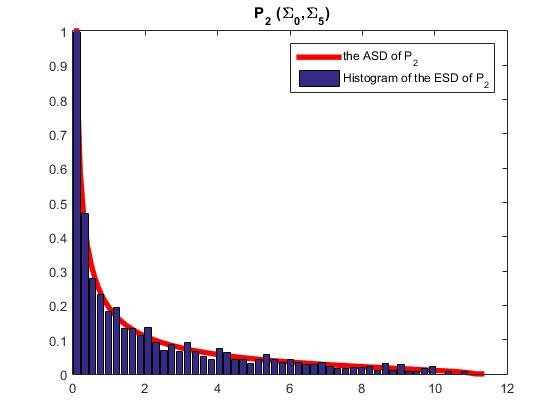}
 }
 \caption{Detection results with the multivariate nonlinear polynomial $P_2$}
 \label{result_p2}
 \end{figure}

Fig.~\ref{result_p2} shows the results. The red curve represents  the ASD of ${P_2}$ obtained through the method in \ref{asd2}. The values of $s$  in $\textbf{C}_1$--$\textbf{C}_5$ are presented in Tab.~\ref{tabs2}.

1) For Fig.~\ref{fig3e}, the histogram agrees perfectly with the theoretical curve  when in the absence of spikes; for other figures (a)-(d), there exist \textit{outliers}.

2) Sort by the values of $s$: $0=\textbf{C}_5<\textbf{C}_2<\textbf{C}_3 \ll \textbf{C}_1 \ll\textbf{C}_4$.

Result 2) show that the step signals and the ramp signals are remarkably distinguished by the size of \textit{outliers}. This implies that the nonlinear polynomial ${P_2}$ is more sensitive to outliers for the same signal-to-noise ratio, compared with its linear case. Furthermore,  these results indicate that  the anomaly's influence on the grid can be estimated quantitatively  by the sizes of \textit{outliers}.

Compared with linearity, nonlinearity is more flexible in problem modeling and closer to the reality. Some other multivariate nonlinear polynomials may be more effective for the power grid with special load characteristics. The search for such an optimal nonlinear polynomial is beyond the scope of this paper.

\begin{table}[h]

\centering

\caption{The values of $s$ in $\textbf{C}_1$--$\textbf{C}_5$}

\begin{tabular}{p{1cm}|p{3.4cm}|p{0.9cm}}
\hline
  Cross Section & Description &$s$\\
\hline
  $\textbf{C}_1$ & Existence of a step signal&0.9027\\
  $\textbf{C}_2$ & Steady load growth&0.1152\\
  $\textbf{C}_3$ & Steady load growth&0.2783\\
  $\textbf{C}_4$ & Chaos due to voltage collapse&5.2332\\
  $\textbf{C}_5$ & No signa &0.000\\
\hline
\end{tabular}
\label{tabs2}
 \end{table}

\subsection{Case 3: Fault Location with Simulation Data }
\label{case3}
In Case 3, the indicator $L_i$ defined in \eqref{indicator} is used to conduct fault locations. As introduced in \ref{faultlocation}, the proposed  fault location method defined in~\eqref{loc} is based on the hypothesis testing.
Case 1 and Case 2 validate our proposed detection method and show that the  nonlinear polynomial  $P_2$ is more effective  than the linear one $P_1.$  Thus, $P_2$ is selected as the test statistics in this case.

\begin{figure}[htbp]
 \centering
 \subfloat[Existence of a step signal for Bus 22]{\label{fig4a}
 \includegraphics[width=0.25\textwidth]{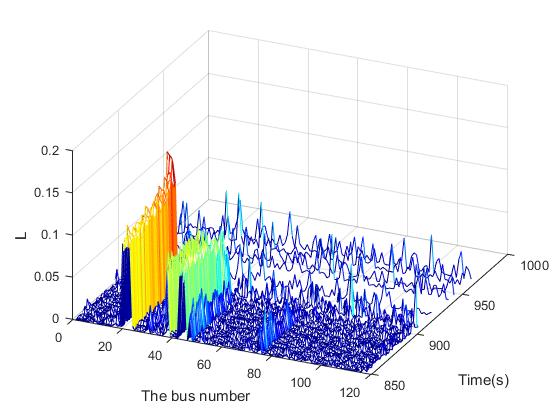}
 }
 \subfloat[Steady load growth for Bus 22]{\label{fig4b}
 \includegraphics[width=0.25\textwidth]{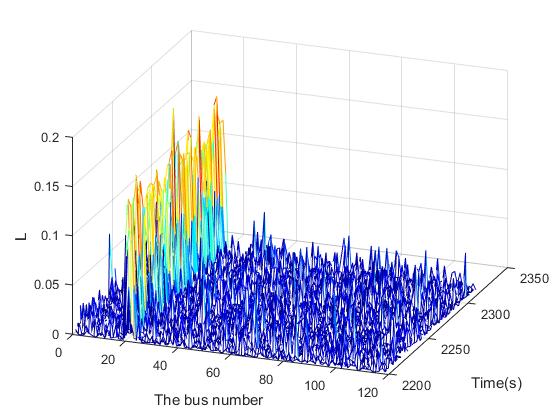}
 }\\
 \subfloat[Steady load growth for Bus 52]{\label{fig4c}
 \includegraphics[width=0.25\textwidth]{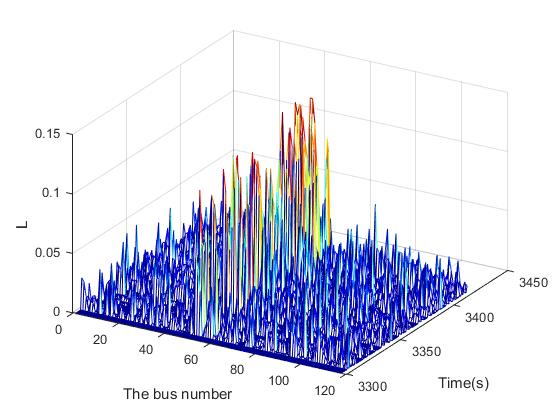}
 }
 \subfloat[Voltage collapse]{\label{fig4d}
 \includegraphics[width=0.25\textwidth]{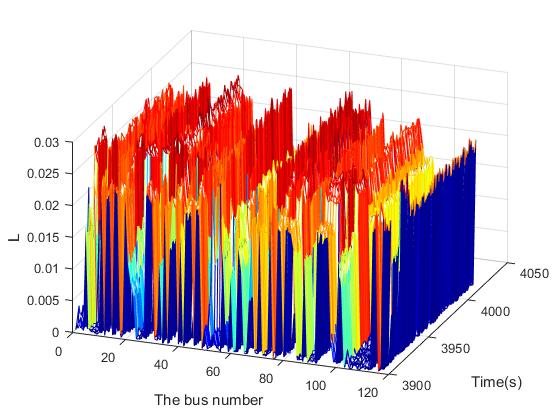}
 }
 \caption{Fault location results using the proposed method in~\ref{faultlocation}.}
 \label{locationfig}
 \end{figure}

Fig.~\ref{locationfig}  is the 3D Plot for the time series of the indicator $L_i.$   Figures~\ref{fig4a},~\ref{fig4c} and~\ref{fig4c} show that the proposed indicator $L_i$ captures the bus information that is most affected by the topology change, and the location results are  consistent with the the event description in Table~\ref{tab1}. The fault location fails in Fig.~\ref{fig4d} due to voltage collapse. This result is close to the real fact.

\begin{figure}[htb]
 \centering
 \subfloat[Existence of a step signal for Bus 22]{\label{Location_lim_c1}
 \includegraphics[width=0.25\textwidth]{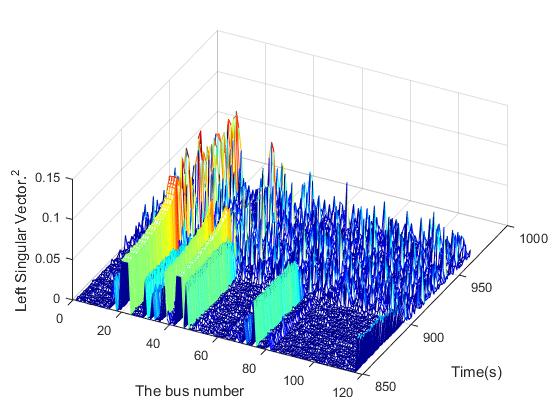}
 }
 \subfloat[Steady load growth for Bus 22]{\label{Location_lim_c2}
 \includegraphics[width=0.25\textwidth]{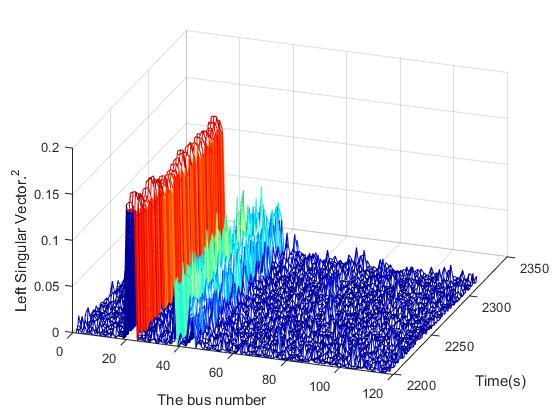}
 }\\
 \subfloat[Steady load growth for Bus 52]{\label{Location_lim_c3}
 \includegraphics[width=0.25\textwidth]{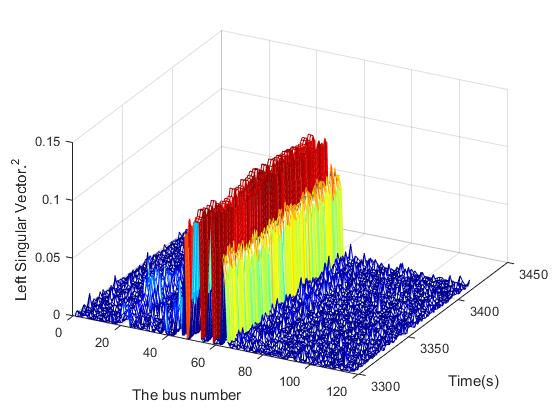}
 }
 \subfloat[Voltage collapse]{\label{Location_lim_c4}
 \includegraphics[width=0.25\textwidth]{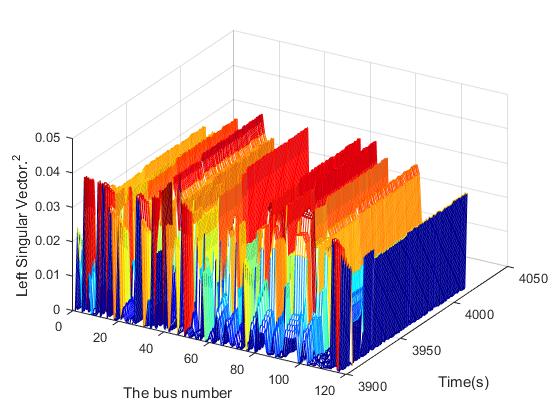}
 }
 \caption{Fault location results only using the largest singular value's corresponding left singular vector}
 \label{locationfiglim}
 \end{figure}

Fig.~\ref{locationfiglim} is the result obtained by Lim's method in~\cite{Lim2013Model}.  The location results are close to our method at most time points. However, in  some points, it is not easy to determine which bus is the most vulnerable to the load variation, especially in Fig.~\ref{Location_lim_c1} and~\ref{Location_lim_c3}.  The reason is that the peaks in other buses are not independent of the statistic information corresponding to the largest singular value.

\begin{figure}[h]
 \centering
 \subfloat[Distribution of the eigenvector components of four different eigenvalues well inside the support]{\label{fig:nonoutliereigv}
 \includegraphics[width=0.35\textwidth]{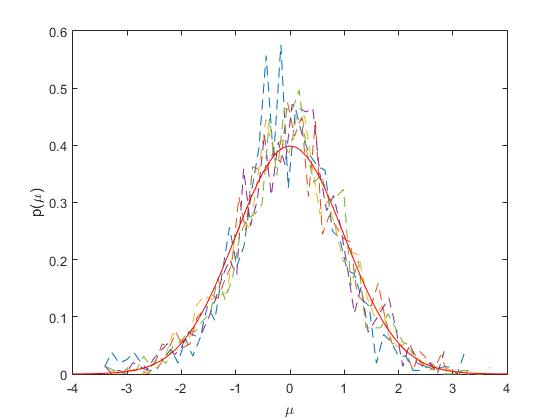}
 }\\
 \subfloat[Distribution of the eigenvector components of \textit{outliers}]{\label{fig:outliereigv}
 \includegraphics[width=0.35\textwidth]{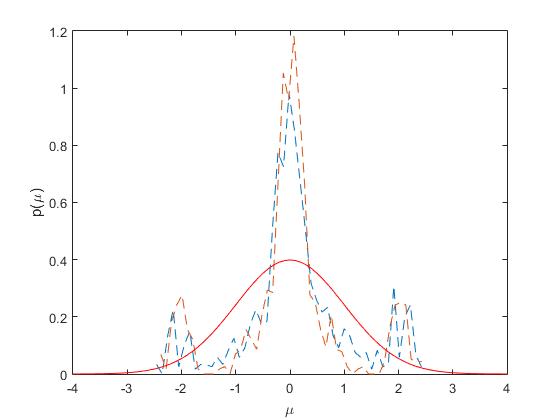}
 }
 \caption{Distribution of the eigenvector components;  the
red solid line represents the standard normal distribution}
 \label{eigv}
 \end{figure}

The reason  why our method performs better is given by studying the distribution of the corresponding eigenvectors. We select the cross section $\textbf{C}_1$ as an example. Let $v_{ik}$ denote the $i$-th component of the eigenvector corresponding to the eigenvalue $\lambda_k$. Then, we normalize it such that $\sum\limits_{i = 1}^N {v_{ik}^2}  = N$.  For a fixed $k$, the distribution of $\mu=v_{ik}$ is denoted by $p(\mu)$.  The distribution of $\mu$ is plotted in Fig.~\ref{eigv}  by dashed lines. The red solid line represents the \textit{standard normal distribution}. As shown in Fig.~\ref{fig:nonoutliereigv}, the $p(\mu)$ for four randomly selected eigenvalues well inside the support  fits extremely well with the \textit{standard normal distribution}. In some sense,  there is no signal contained in these eigenvectors. On the contrary, the $p(\mu)$ for \textit{outliers} is markedly different from the \textit{standard normal distribution} in Fig.~\ref{fig:outliereigv}. This means that not only the largest eigenvalue but also all other \textit{outliers} contain the most statistical information about the signal. In other words, all outliers matter!

\subsection{Case 4: Fault Location with Real 34-PMU Data}
\label{case4}
In this subsection, we evaluate the fault location indicator $L_i$ with real-world 34-PMU data. The real power data is a chain-reaction fault that happened in 2013 in one large power grid in China. The sample rate is 50 \textit{Hz} and the total sample time is $284$ seconds (s). Fig.~\ref{34-PMU_powerflow} and Fig.~\ref{34-PMU_chain-reaction}  illustrate the three-dimensional power flow at the whole time and the fault time respectively. The chain-reaction fault starts at $t=65.4$s.
\begin{figure}[htb]
 \centering
\subfloat[The realistic 34-PMU power flow.]{\label{34PMU_powerflow}
 \includegraphics[width=0.3\textwidth]{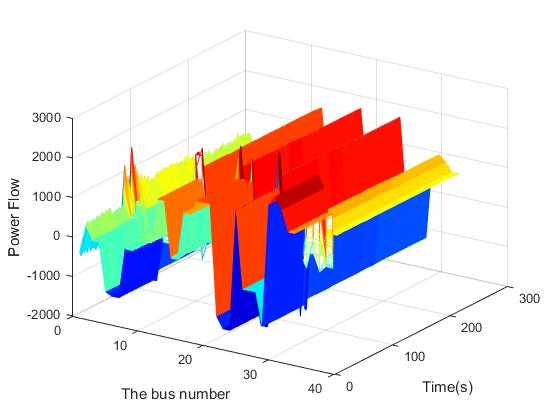}
 }\\
 \subfloat[The realistic 34-PMU power flow around the chain-reaction fault occurrence.]{\label{34PMU_chain-reaction}
 \includegraphics[width=0.3\textwidth]{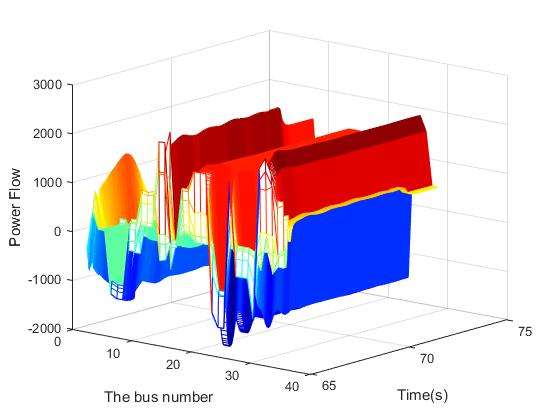}
 }
\caption{3D Plot for time series of the 34-PMU power flow}
 \label{34-PMU}
 \end{figure}

\begin{figure}[htpb]
\centering
\begin{overpic}[scale=0.35
]{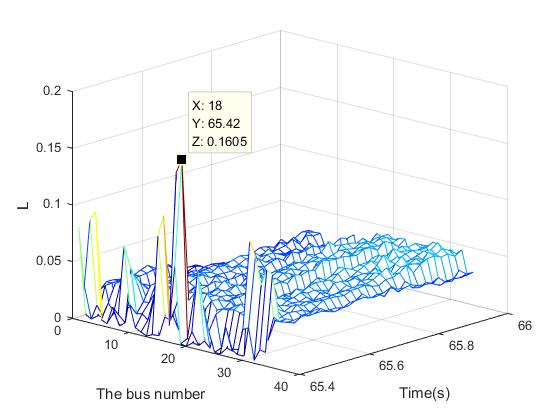}
\end{overpic}
\caption{Fault location result with the 34-PMU data}
\label{fig:34-PMU_location}
\end{figure}

Similarly to the data processing in simulation Case 3, set the sample dimension $N=34$ and the sample length $T=N$. The location of the most sensitive bus can be determined using $L_i$ defined in~\eqref{indicator}, using the method of~\eqref{loc}. The result shown in Fig.~\ref{fig:34-PMU_location} illustrates that the 18-th PMU ($X=18$) is the most sensitive one which is in agreement with the actual accident situation. This case validates the proposed method in real-life grid.

\section{The estimation of the signal strength}
\label{signalest}
Here we estimate the signal strength under the linear assumption:
\begin{equation}
\label{eqmoedl}
V = A+N
\end{equation}
where $V$ is the grid data matrix, $N$ is the noise matrix and $A$ represents the signal.
Let $V_{(n,i)}$ denotes the $i$-th $n\times n$ sampling matrix and $V_i=N_{n,i} + A_{n,i}$. Define the matrix product as
\begin{equation}
  {M_i}: =V_0V_i=N_0(N_i+A_i)
 \end{equation}
 where $i\geq1$.

It is natural to wonder how close the eigenvalues of $M_i$  are  to those of $A_i$.
\newtheorem{theoremdis}{\textbf{Theorem}}[section]
 \begin{theorem}[\cite{Coston2017Outliers}]
Let $m\geq1$ be an integer, and assume $\xi_1,...,\xi_m$ are complex-valued random variables. For each  $n\geq1$, let $N_{n,1},...,N_{n,m}$ be an $n\times n$ i.i.d random matrix with atom variable  $\xi_1,...,\xi_m$, respectively. In addition, for each $1\leq k\leq m$, let $A_{n,k}$ be a deterministic $n\times n$ matrix with rank $\rm O(1)$ and operator norm $\rm O(1)$. Define the products
\begin{equation}
\label{eqdis}
{M_n}: = \prod\limits_{k = 1}^m {(\frac{1}{{\sqrt n }}{N_{n,k}} + {A_{n,k}})},  A_n:=\prod\limits_{k = 1}^m {{A_{n,k}}}
\end{equation}
and $\sigma:=\sigma_1\cdot\cdot\cdot \sigma_m$. Let $\varepsilon>0$, and suppose that for all sufficiently large $n$, there are no eigenvalues of $A_n$ in the band $\{ z\in \mathbb{C}:\sigma +\varepsilon < |z| <\sigma +3\varepsilon \}$, and there are $j$ eigenvalues $\lambda_1(M_n),...,\lambda_1(M_n)$ of the product $P_n$ in the region $\{z\in \mathbb{C}: |z| \geq\sigma +2\varepsilon  \}$, and after labeling these eigenvalues properly,
\begin{equation}
\label{eqdisth}
 \lambda {}_i({M_n}) = \lambda {}_i({A_n}) + o(1)
 \end{equation}
as $n\longrightarrow\infty$ for each $1\leq i\leq j$.

  \label{thmdiscussion}
  \end{theorem}

Theorem~\ref{thmdiscussion} reveals two main points:
 \begin{enumerate}[$\bullet$]
\item when the sizes of matrices are large, the cross terms in \eqref{eqdis} can be negligible outliers exit if signals exist in the system.
 \item  the combined strength of the signals can be bounded according to \eqref{eqdisth}.
\end{enumerate}
The outliers of $M_n$ are asymptotically close to the outliers of the product $A_n$ defined in~\eqref{eqdis}. This implies that the combined signal strength can be estimated by calculating the eigenvalues of $M_i$ directly. In practice, the $M_i$ can be obtained by measurements but the $A_n$ is difficult to know.

\begin{figure}[htbp]
 \subfloat[Step signal $V_1$]{\label{circlea}
 \includegraphics[width=0.25\textwidth]{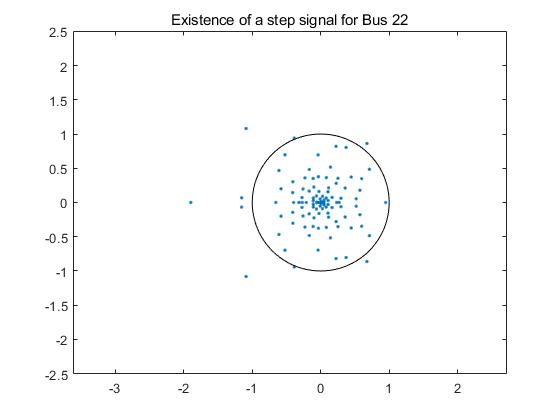}
 }
 \subfloat[Stable growth A $V_2$]{\label{circleb}
 \includegraphics[width=0.25\textwidth]{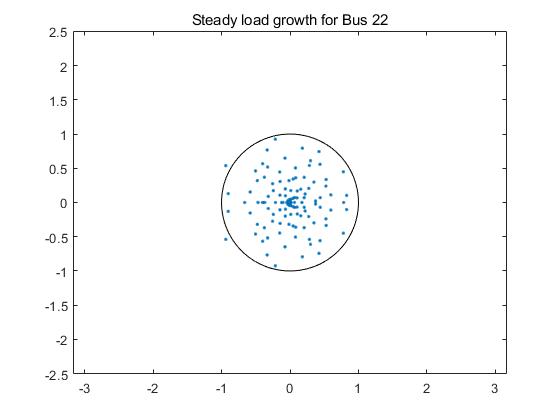}
 }\\
 \subfloat[Stable growth B $V_3$]{\label{circlec}
 \includegraphics[width=0.25\textwidth]{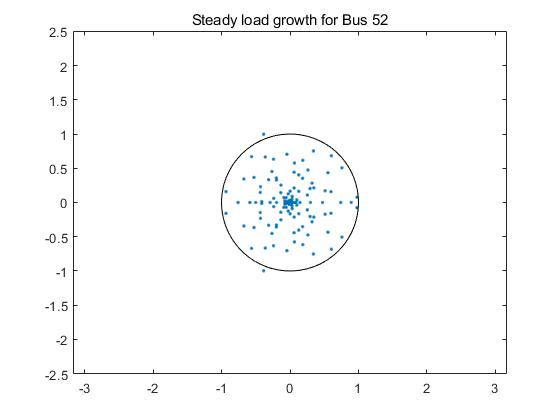}
 }
 \subfloat[Voltage collapse $V_4$]{\label{circled}
 \includegraphics[width=0.25\textwidth]{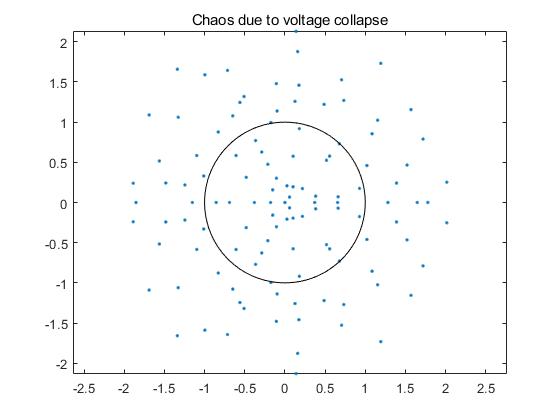}
 }\\
 \subfloat[White noises $V_5$]{\label{circlee}
 \includegraphics[width=0.25\textwidth]{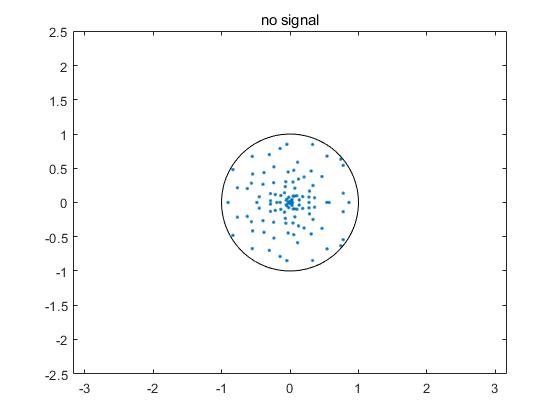}
 }
 \caption{The eigenvalues of $V_0V_i$}
\label{result_circle}
 \end{figure}

We illustrate our approach using the simulated data used in Section~\ref{CASE STUDIES}.
The eigenvalues of $V_0V_i, i=1,2,3,4,5$ are plotted in Fig~\ref{result_circle}, respectively.

The results  in Fig~\ref{result_circle} show an interesting phenomenon of products of random matrices: outliers appear only in the presence of the signals. 

\section{CONCLUSION}
\label{CONCLUSION}
Built upon random matrix theory (RMT), we obtain new statistical models using massive datasets across the power grid. In this paper, we take advantage of a breakthrough by~\cite{Speicher2015Polynomials,Belinschi2013Analytic} in free probability to calculate the asymptotic spectrum distribution of the self-adjoint matrix polynomials. Our problem of anomaly detection is formulated in terms of hypothesis testing.   Fault location is also conducted. The new approach has advantages over previous ones. The results generated from the 2383-bus system agree with the asymptotic theory much better than those  the 118-bus system.

As a starting point, this paper considers only two simplest examples of the self-adjoint matrix polynomials: ${P_1}({\Sigma _0},{\Sigma _1}) = {\Sigma _1} - {\Sigma _0}$ and ${P_2}({\Sigma _0},{\Sigma _1}) = {({\Sigma _1} - {\Sigma _0})^2},$ where ${\Sigma _0}$ and ${\Sigma _1}$ are the large-dimensional sample covariance matrices, respectively, for the null and alternative hypotheses. This problem is related to the difference between two mixed quantum states (e.g., Wishart matrices with fixed trace)~\cite{mejia2016difference,puchala2016distinguishability,nechita2016almost} where analysis is conducted by free probability.  Specifically, if we define the mixed quantum states as ${\rho _0} = {\Sigma _0}/\operatorname{Tr} \left( {{\Sigma _0}} \right),{\rho _1} = {\Sigma _1}/\operatorname{Tr} \left( {{\Sigma _1}} \right),$ we can study the difference $p{\rho _1} - q{\rho _0},{\text{ for }}p,q \in \mathbb{R},$ following~\cite{mejia2016difference}.

Different trace functions as done in~\cite{puchala2016distinguishability} can be considered in the future. Specifically, we consider the linear eigenvalue statistics (LES) of the difference of the two quantum states $\operatorname{Tr} f\left( {{\rho _1} - {\rho _0}} \right) = \sum\limits_{i = 1}^n {f\left( {{\lambda _i}} \right)} ,$ where $f:\mathbb{R} \to \mathbb{R}$ is an arbitrary function of some certain smooth properties and ${{\lambda _i}},i=1,...,n$ is the  $i-$th eigenvalue of the difference ${{\rho _1} - {\rho _0}}.$ This is the extension of the LES for one single random matrix in~\cite{He2016Designing}. What is the optimal function $f$?

The algorithm of~\cite{Speicher2015Polynomials,Belinschi2013Analytic} is very general. The techniques in quantum information theory, e.g.,~\cite{mejia2016difference,puchala2016distinguishability,nechita2016almost}, have more explicit expressions to give use more transparent solutions to our problems at hand. All these papers are unified within the paradigm of free probability, a fast growing branch of random matrix theory. Note that the first use of RMT in a large power grid was by the same authors of this paper  in~\cite{he2015arch}. The whole paradigm of using RMT allows one to exploit the theory of \textit{asymptotically large} random matrices. The whole framework lies in the empirical observation that the asymptotic limits are very close to that finite-size random matrices, even for \textit{moderate sizes}! The empirical success of using the asymptotic theory of~\cite{Speicher2015Polynomials,Belinschi2013Analytic} in finite-size cases of our studies in a power grid will pave the way for studies of big data data analytics using other massive datasets collected in such as internet of things (IOT).

In this paper, the simple white noise representations are adopted to model the  stochastic variation in load; some recent studies have adopted Ornstein-Uhlenbeck process~\cite{Roberts2016Validation} - Validation of the Ornstein-Uhlenbeck process for load modeling based on PMU measurements. These inspire us to adopt some achievements about Ornstein-Uhlenbeck Process based on RMT in the future work.

\begin{appendices}
\section{The Specific Procedures of calculating The ASD of $P_2$}
\label{appendixA}
The following steps give the precise statement of the algorithm in ~\ref{asd2} .

\begin{enumerate}[step 1]

\item Compute the linearization of $P_2$
\[{{L}_{P_2}}={c}\otimes 1+{{b}_{0}}\otimes {{\Sigma}_{0}}+ {{b}_{1}}\otimes {{\Sigma}_{1}}\] through Anderson's linearization trick.

\item Compute the  Cauchy transform ${{G}_{{{b}_{j}}\otimes {{\Sigma}_{j}}}}(b)$ through the scalar-valued Cauchy transforms :
\[{{G}_{{{b}_{j}}\otimes {{\Sigma}_{j}}}}(b)=\underset{\varepsilon \to 0}{\mathop{\lim }}\,-\frac{1}{\pi }\int_{\mathbb{R}}{(b-t{{b}_{j}}}{{)}^{-1}}\Im ({{G}_{{{\Sigma}_{j}}}}(t+i\varepsilon ))dt.\]
for $j=0,1$.
\item Calculate the Cauchy transform of
\[{{L}_{P_2}}-{c}\otimes 1={{b}_{0}}\otimes {{\Sigma}_{0}}+ {{b}_{1}}\otimes {{\Sigma}_{1}}\]
by applying Theorem \ref{th1}.
The Cauchy transform of ${{L}_{P_2}}$ is then given by \[{{G}_{{{L}_{P_2}}}}(b)={{G}_{{{L}_{P_2}}-{c}\otimes 1}}(b-{c}).\]

\item According to  Corollary  \ref{co1}, the scalar-valued Cauchy transform ${{G}_{P_2}}(z)$ of $P_2$ 	is obtained by
\[{{G}_{P_2}}(z)\text{=}\underset{\varepsilon \to \text{0}}{\mathop{\lim }}\,{{\left[ G{}_{{{L}_{P_2}}}({{\Lambda }_{\varepsilon }}(z)) \right]}_{1,1}}\quad for\; all \;z\in {{\mathbb{C}}^{+}}.\]

\item Compute the distribution of $P_2$ via the Stieltjes inversion formula.

\end{enumerate}

\section{The standard IEEE 118-bus system and The load variation}
\label{appendixB}

\begin{figure}[h]
\centering
\begin{overpic}[scale=0.2
]{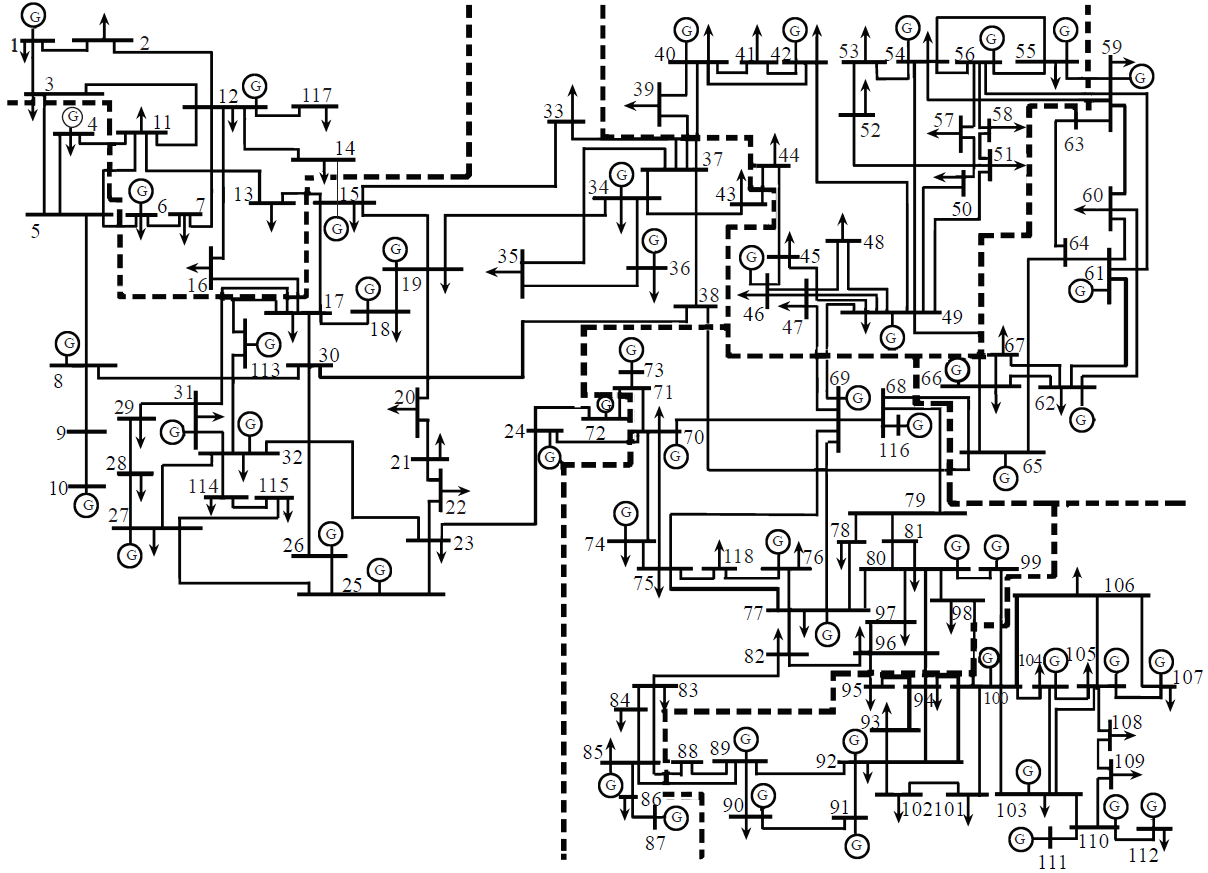}
\end{overpic}
\caption{The network structure for the IEEE 118-bus system.}
\label{fig:IEEE118network}
\end{figure}

\begin{figure}[htbp]
\centering
\begin{overpic}[scale=0.2
]{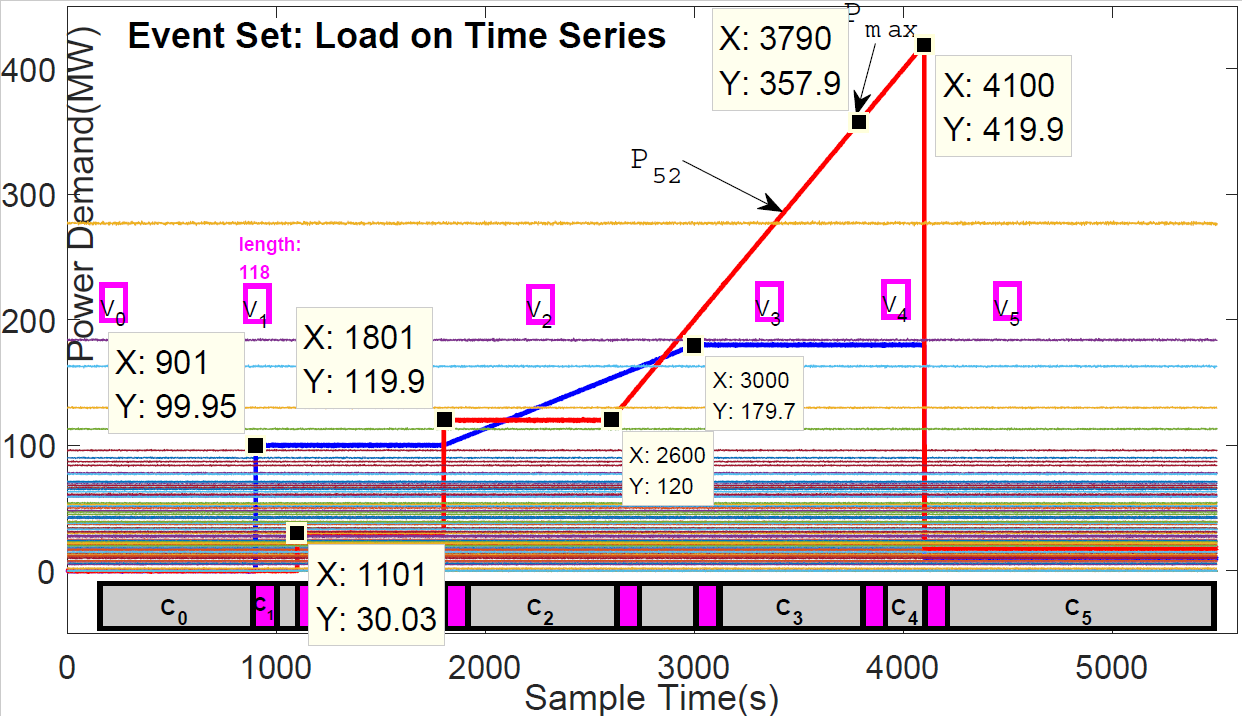}

\end{overpic}
\caption{The event assumptions on time series.}
\label{fig:loadevent}
\end{figure}

\section{The 2383-Bus case}
\label{appendixC}

\begin{figure*}[htb]
 \centering
 \subfloat[Steady load growth for Bus 59]{\label{fig7a}
 \includegraphics[width=0.35\textwidth]{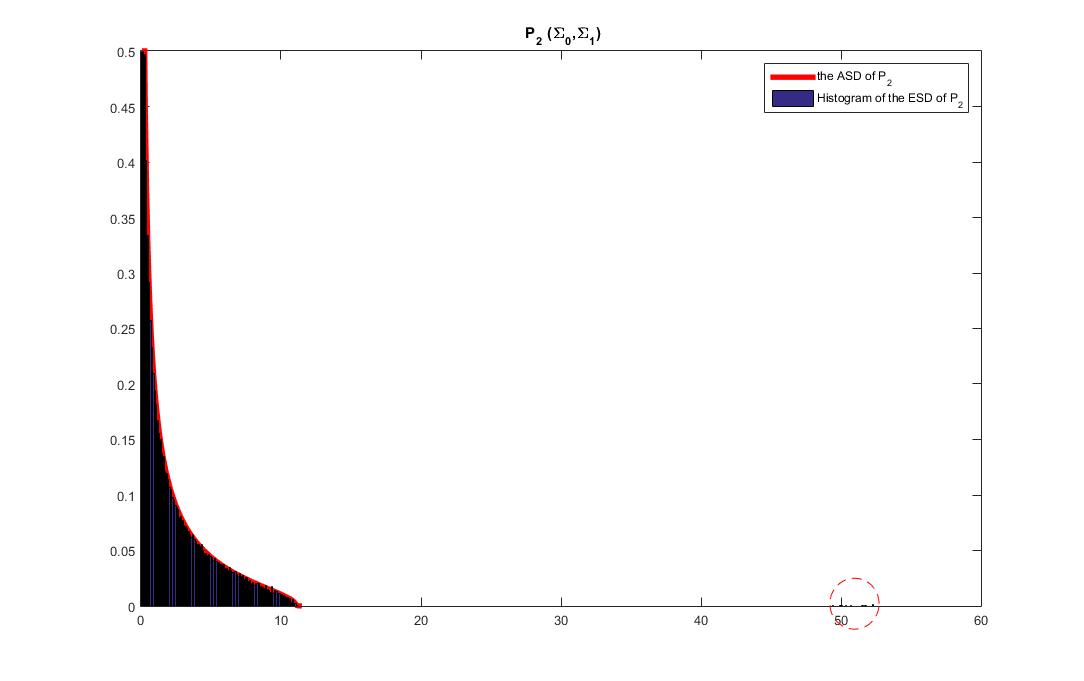}
 }
 \subfloat[Existence of a step signal for Bus 59]{\label{fig7b}
 \includegraphics[width=0.35\textwidth]{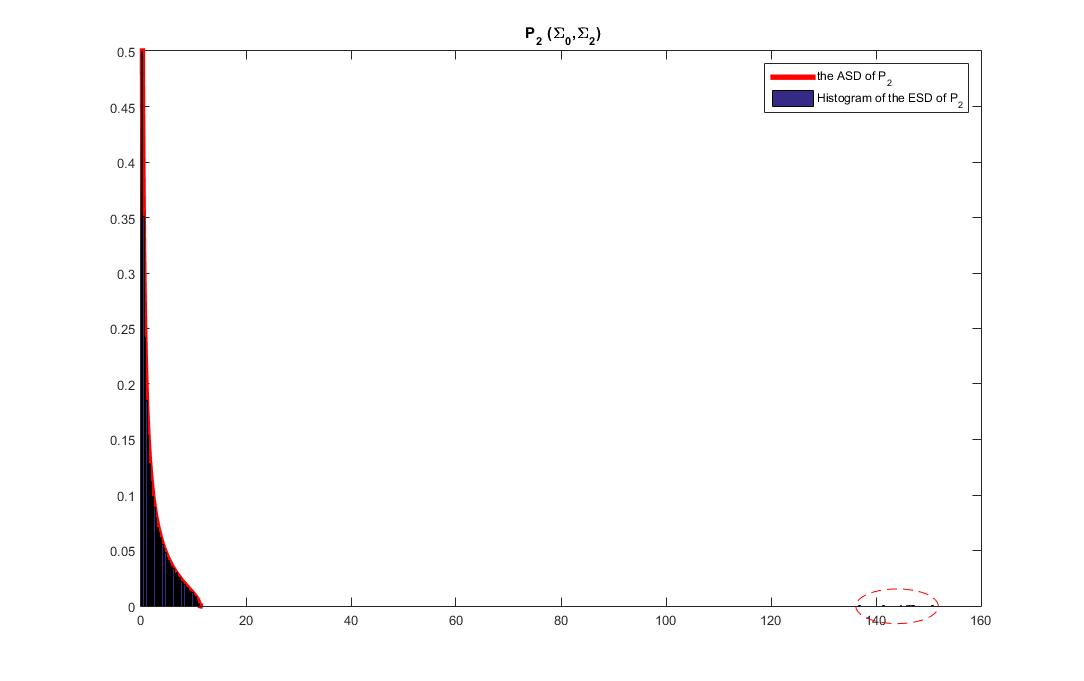}
 }
 \subfloat[Voltage collapse]{\label{fig7c}
 \includegraphics[width=0.35\textwidth]{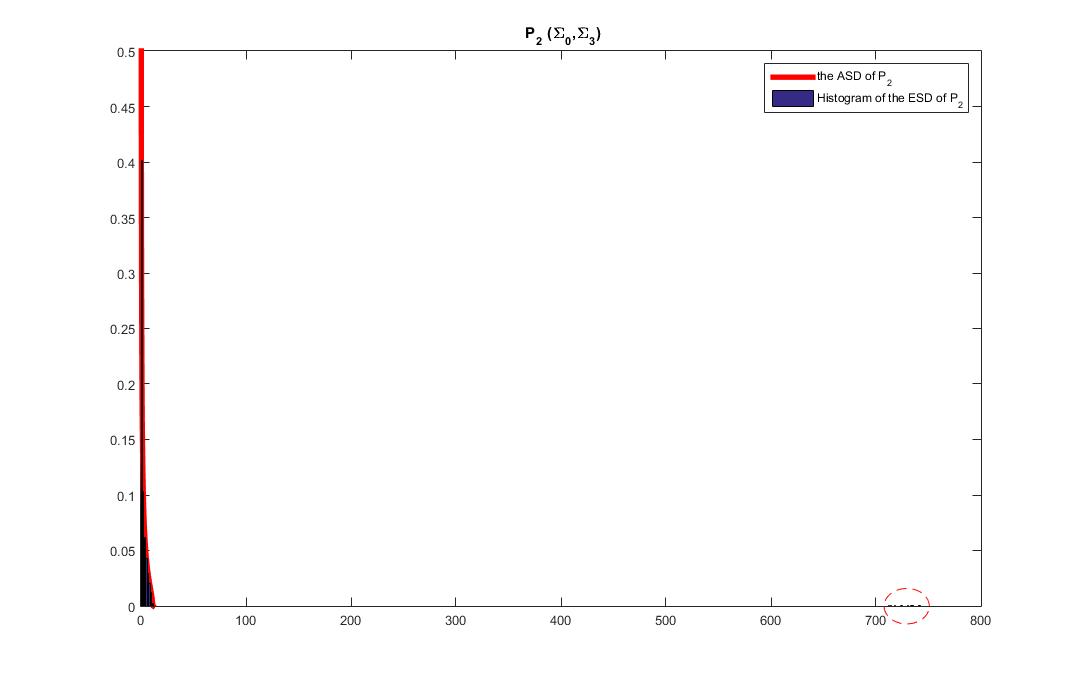}
 }
 \caption{Detection results of the 2383-bus case}
 \label{detectionfig2383}
 \end{figure*}

\begin{figure*}[htb]
 \centering
 \subfloat[Steady load growth for Bus 59]{\label{fig8a}
 \includegraphics[width=0.35\textwidth]{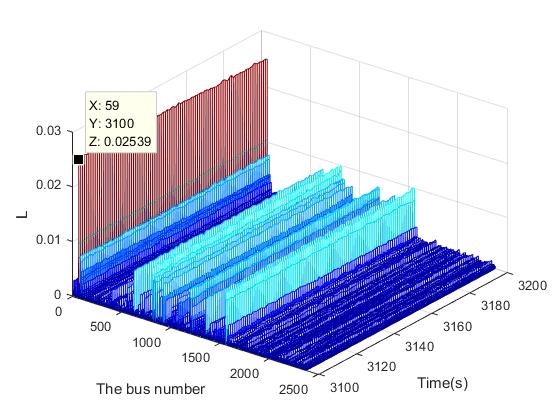}
 }
 \subfloat[Existence of a step signal for Bus 59]{\label{fig8b}
 \includegraphics[width=0.35\textwidth]{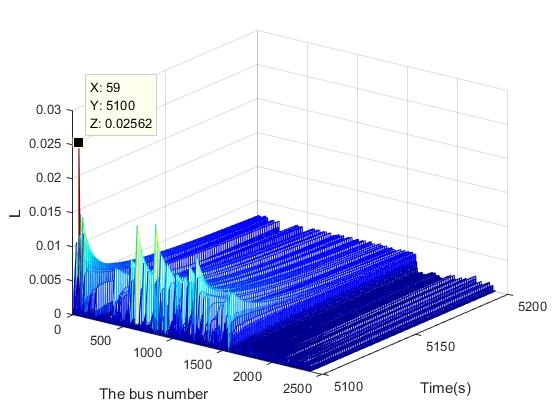}
 }
 \subfloat[Voltage collapse]{\label{fig8c}
 \includegraphics[width=0.35\textwidth]{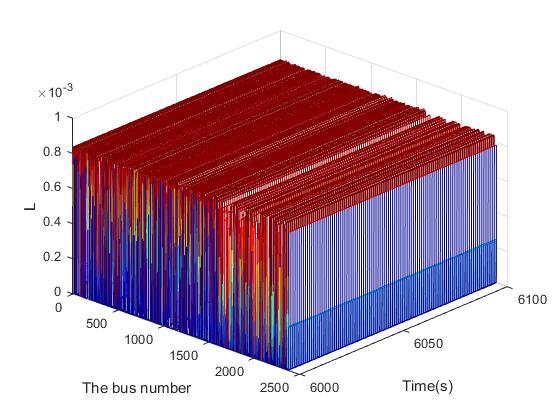}
 }
 \caption{Fault location results of the 2383-bus case}
 \label{locationfig2383}
 \end{figure*}

For simplicity, the signals for each bus are shown in Tab.~\ref{tab_2383}. The detection results, the values of $s$ location results are shown in Fig.\ref{detectionfig2383}, Tab.\ref{tabs3} and
Fig.~\ref{locationfig2383} respectively.
The results generated from the 2383-bus system are the same as  the 118-bus system.
\begin{table}[h]
\centering
\caption{Descriptions of the 2383-bus system status}
\begin{tabular}{p{1cm}|p{1.5cm}|p{3.4cm}}
\hline
  Bus & Duration(s) &Descripiton\\
\hline
  59 & 3100 $\sim$ 3200&Steady load growth \\
     & 5100 &Existence of a step signal\\
     & 6000 $\sim$ 6100 &Chaos due to voltage collapse\\
\hline
 Others & 1 $\sim$ 10000&No signal\\
\hline
\end{tabular}
\label{tab_2383}
\end{table}

\begin{table}[h]
\centering
\caption{The values of $s$ }
\begin{tabular}{p{1.5cm}|p{3.4cm}|p{0.9cm}}
\hline
  Durations & Description &$s$\\
\hline
3100 $\sim$ 3200&Steady load growth& 0.2538 \\
 5100 &Existence of a step signal& 2.045\\
 6000 $\sim$ 6100 &Chaos due to voltage collapse&9.668\\
\hline
\end{tabular}
\label{tabs3}
 \end{table}

\end{appendices}

\bibliographystyle{IEEEtran}
\bibliography{lznbib}

\end{document}